# Digital Speech Algorithms for Speaker De-Identification

Stefano Marinozzi
Institute for Electronic Engeenering
Università Politecnica delle Marche
Ancona, Italy
stefano.marinozzi87@gmail.com

Marcos Faundez-Zanuy
EUPMt , Fundació
Tecnocampus
Mataró, Spain
faundez@tecnocampus.cat

*Abstract*— **The present work is based on the COST Action IC1206 for De-identification in multimedia content [1]. It was performed to test four algorithms of voice modifications on a speech gender recognizer to find the degree of modification of pitch when the speech recognizer have the probability of success equal to the probability of failure. The purpose of this analysis is to assess the intensity of the speech tone modification, the quality, the reversibility and not-reversibility of the changes made.**

*Keywords—De-Identification;Speech Algorithms*

## I. INTRODUCTION

In recent times many useful service have become available via web or over the telephone. With the increased usage of such services, the users are also becoming more aware of privacy implication of their use. Therefore application that can assure that users can protect their privacy are becoming more attractive. Methods concerning person de-identification in still imager or video have already been proposed [2], and try to mask identification features such as faces, silhouettes, posture, etch.

There is also a need for de-identification technologies in voice driven application. For example, conversation may be recorded in call centers for various purposes, such analysis of operator mistakes, making the communication protocol more efficient in general or to prove a call has actually been made in case of complaints etc. In many cases, the identity of the caller is not important for given purpose, and customers may legitimately wonder why it should be recorded and kept.
In [3], [4] voice transformation were implemented to de-identify a small set of speaker and tested with automatic speaker identification system. Voice transformation was successful in concealing identities of source speakers against the GMM-based speaker identification system. However in those experiments, speech samples from each person to be de-identified had to be available in advance in order to estimate the transformation parameters. In addition, those samples were parallel utterance, with the same text spoken by source and target speakers. Also to de-identify a speaker, his identity has to be known first, so that his corresponding voice transformation can be used for de-identification. This may also be a limitation in some cases, where the user doesn't want to identify with the systems at all.

In a scenario with a closed set of speakers to be de-identified, this may be acceptable, but for application it would not be practical. In that case, the number of potential users of the system is extremely large, and many users will only use the system once. A requirement that the users has to supply would be inconvenient. For a practical application in such systems, it would be desirable that any new user can use the system immediately, without having to enroll with the system first in any way or identify himself, even for the purpose of de-identification.

In this paper we propose two kinds of principal approaches for voice modification: the first one is based on the Dolson and Laroche Phase Vocoder [5], the second one is based the Vocal Tract Length Normalization (VTLN) based on frequency warping factor [6], [7]. These approaches are focused on modifying the non-linguistic characteristic of a given utterances without affecting its textual contents and, so, to preserve the intelligibility of the spoken language. The state-of-the-art of gender recognition system nearly gives a perfect recognition performance in a clean condition and using normal voices [8]. Using the threshold from this condition, we applied 25 steps of pitch modifications at all the BiosecureID database files [9] (15200 speech files), then, analyzing the response of the gender recognizer with the comportments of the pitch modification algorithms for the masculine and feminine speech, we created the graphics that will be discussed. Results show that the probability of success changes depend on the kind of pitch modification algorithms and the gender of the speaker.
To evaluate the quality of different algorithms, was performed a listening test with 15 candidates, and was also compared machine performance with human performance.

## II. BRIEF DESCRIPTION OF THE ALGORITHMS

### A. The Phase Vocoder

The Phase Vocoder is presented as a high-quality solution for time-scale modification of signals, pitch-scale modifications usually being implemented as a combination of time scaling and sampling rate conversion. Pitch-shifting using the STFT representation of an audio signal as proposed in [10] is performed in four steps:

1. Peak detection: The simplest scheme consists of declaring that a bin is a peak if its magnitude is larger than that of its two (or four) nearest neighbors. It is assumed that each detected peak represents a sinusoidal component.

2. Define regions of influence: The region of influence is the sub-band around a spectral peak in which it is assumed that all phase values are dominated by the peak's phase. The boundaries of these sub-bands can be defined halfway between two peaks or at the lowest magnitude bin between two peaks.

3. Coefficient shift: Peaks and their regions of influence are shifted by frequency $\Delta\omega_m$, where m is the peak index. If the relative amplitudes and phases of the bins around a sinusoidal peak are preserved during the translation, then the time-domain signal corresponding to the shifted peak is simply a sinusoid at a different frequency, modulated by the same analysis window.

4. Phase propagation: Since the frequencies of underlying sinusoids have been changed during the coefficient shift, phase-coherence from one frame to the next is lost. To avoid arte-facts due to inter-frame phase inconsistency, phase values need to be updated.

The difference between the two kinds of Phase Vocoder used is in the function that describe the Phase propagation [11].

### B. *The Vocal Tract Length Normalization (VTLN)*

In order to change the voice into another, the spectrum of a frame has to be transformed [12]. Vocal tract length normalization is used to warp the spectrum of a frame, i.e., stretch or compress the spectrum with respect to the frequency axis which represents normalized frequencies in the range $0 \leq \omega \leq \pi$.

Frequencies are altered according to a warping function $g(\omega)$ which needs to be a monotonous function returning values between 0 and $\pi$. g returns the warped position of the original frequency. Commonly, g depends on a warping parameter affecting the shape of the function. The warping function is one of five predefined functions as shown in Table I. *Fig. 1* shows the corresponding illustration of these functions.

TABLE I. WARPING FUNCTION

| Type | Formula |
|---|---|
| Symmetric | $g(\omega,\alpha) = \begin{cases} \alpha\omega_1 & \omega \leq \omega_0 \\ \alpha\omega_o + \dfrac{\pi - \alpha\omega_0}{\pi - \omega_0}(\omega - \omega_0), & \omega > \omega_0 \end{cases}$ <br> $\omega_0 = \begin{cases} \dfrac{7\pi}{8} & \alpha \leq 1 \\ \dfrac{7\pi}{8\alpha} & \alpha > 1 \end{cases}$ |
| Asymmetric | $g(\omega,\alpha) = \begin{cases} \alpha\omega_1 & \omega \leq \omega_0 \\ \alpha\omega_o + \dfrac{\pi - \alpha\omega_0}{\pi - \omega_0}(\omega - \omega_0), & \omega > \omega_0 \end{cases}$ <br> $\omega_0 = \dfrac{7\pi}{8}$ |
| Quadratic | $g(\omega,\alpha) = \omega + \alpha\left(\left(\dfrac{\omega}{\pi}\right) - \left(\dfrac{\omega}{\pi}\right)^2\right)$ |
| Power | $g(\omega,\alpha) = \pi\left(\dfrac{\omega}{\pi}\right)^\alpha$ |
| Bilinear | $g(\omega,\alpha) = \left\|-\varepsilon\dfrac{z-\alpha}{1-\alpha z}\right\|$ <br> $z = e^{j\omega}$ |

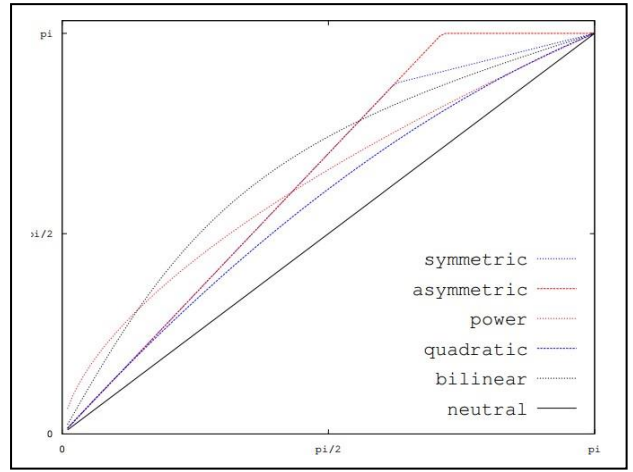

Fig. 1. Warping functions. The warping factor is 0.4 for the bilinear and 0.6 for the power function and 1.4 in all other cases

### III. EXPERIMENTAL RESULTS

We used speech from the BiosecureID database on an ASR trained whit normal voice and tested from disguised voice. The voice modification applied using the Phase Vocoder is equivalent at the increase or decrease of the tone of speech signals, the first one is called VOC and the second one is called VOCF while, for the VTLN it's different and depends from the two warping factor. For the Bilinear's warp function are applied incremental steps of 0.0065 for female speech and decremental steps of 0.0043 for masculine speech; while, for the Quadratic's warp function are applied incremental steps of 0.057 for female speech and decremental steps of 0.029 for masculine speech. *Fig. 2* and *Fig. 3* shows the corresponding trends.

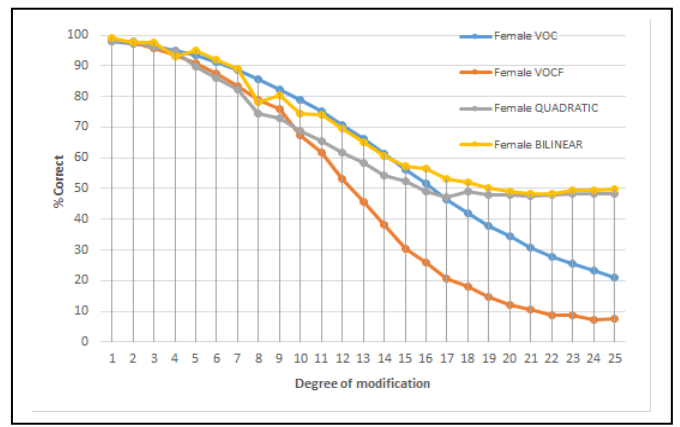

Fig. 2. Results for female voice modification

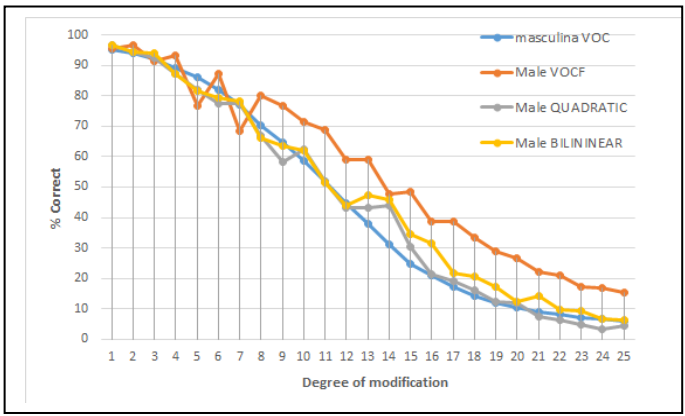

Fig. 3. Results for male voice modification

## IV. HUMAN EVALUATION

De-identification for not revealing the speaker's gender is only useful if the content of the transmitted information is still understandable for human beings. The changes introduced by the algorithms can introduce noise and distortion that gave some concern about the quality of the converted speech that become less natural and intelligible. Consequently, we conducted a human evaluation to investigate the understandability and to verify the gender of the speaker.

Our first test was on speaker identity. For every algorithm, we have made 4 incremental change (equivalents to the degree of modification 7, 13, 19, 25) for a total of 256 audio files.

We asked our listeners to identify if the speaker was male or female. The identification by a human listening is deeper compared to the gender recognition system: the results are similar for all the algorithms that we have used and while for the gender recognition system there is a general decay when the modification of the voice is stronger, the human identification requires a louder change.

*Fig. 4*, *Fig. 5*, *Fig. 6* and *Fig. 7* shows the trends for the four algorithms.

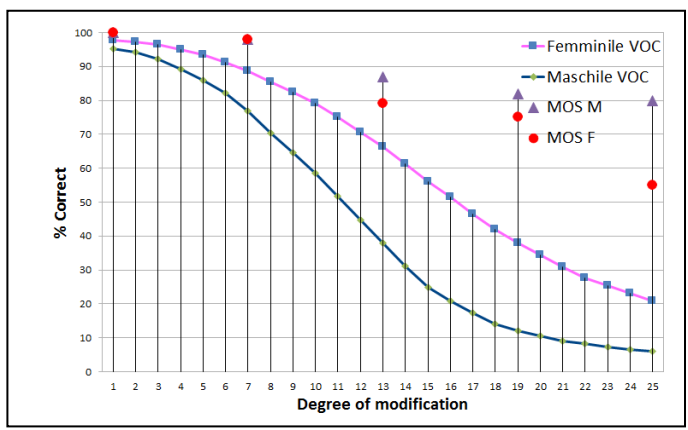

Fig. 4. MOS for VOC algorithm

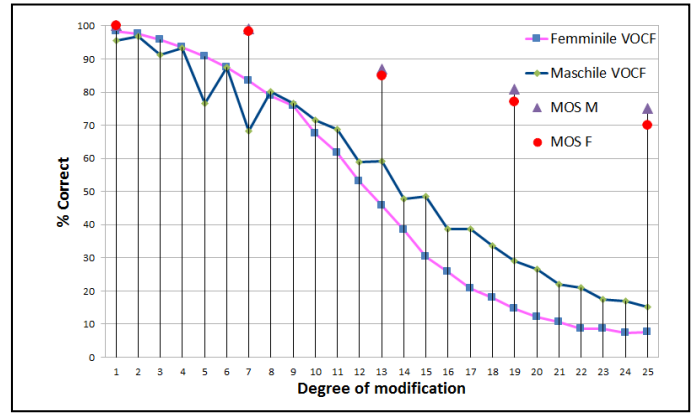

Fig. 5. MOS for VOCF algorithm

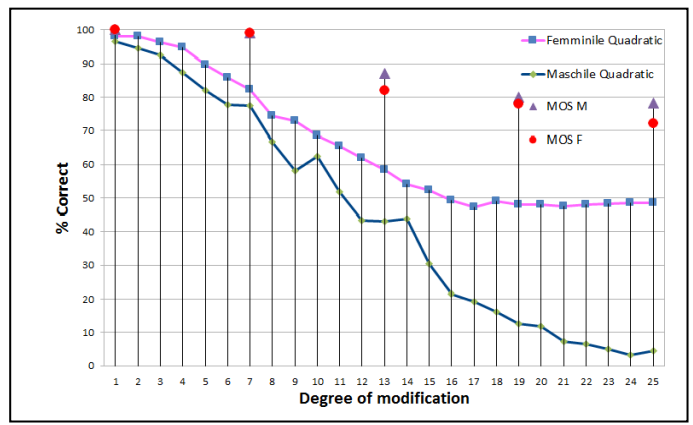

Fig. 6. MOS for QUADRATIC algorithm

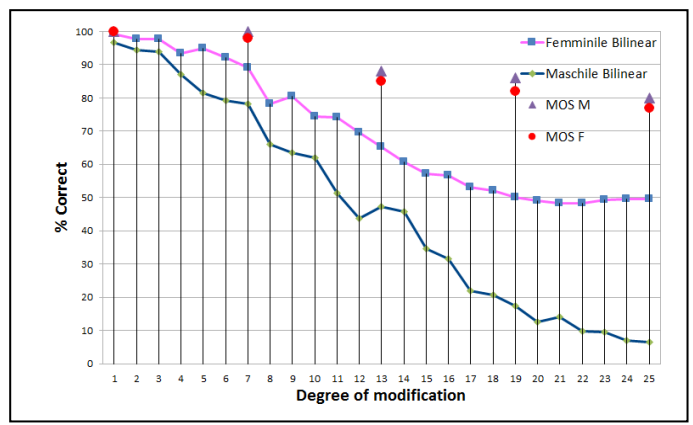

Fig. 7. MOS for BILINEAR algorithm

The trends of the human evaluation are similar for all the kind of algorithms.

The second test was on intelligibility. A successful de-identification process should preserve the understandability of the transmitted content. We played examples of the de-identified speech to listeners and asked them the effort required to understand the meanings of sentences [13]:

5 Complete relaxation possible; no effort required.
4 Attention necessary; no appreciable effort required.
3 Moderate effort required.
2 Considerable effort required.
1 No meaning understood with any feasible effort.

The results obtained are the following:
- Phase Vocoder (VOC): 3,7
- Phase Vocoder (VOCF): 3,3
- VTLN Quadratic: 3,4
- VTLN Bilinear: 3

## V. CONCLUSION AND FUTURE WORK

In this paper, we studied the potential of voice transformation for gender de-identification. The method does not require enrolment of speakers for de-identification thus greatly extending possible applications of the system. We explored different voice transformation strategies including two kinds of Phase Vocoder that permit the reversibility of the changes made with a good intelligibility, and two functions of the Vocal Tract Length Normalization that not permit the reversibility of change made, keeping a sufficient intelligibility.

For the Phase Vocoder (VOC) we can see that when the error rate is at 50%, the degree of modification is 11 for male and 16 for female, while, for the Phase Vocoder (VOCF) the degree of modification is 13 for male and 16 for female.

For the VTLN with the Quadratic's warp function, the degree of modification is 11 for male and 16 for female, while, for the Bilinear's warp function, is 11 for male and 19 for female.

Further experiments were designed to test the ability of humans to recognize unknown speaker. In a carefully controlled experiment, human performance was measured and was compared to the gender recognition system. Results showed that machine performance is not comparable to average human performance.

We hope to expand this work in Automatic Speech Recognition research for children. The scope is to decrease the error rate using models trained on adult speech through the reduction of the pitch of the children speech.

ACKNOWLEDGMENT

This work has been supported by FEDER and Ministerio de ciencia e Innovación, TEC2012-38630-C04-03, COST IC-1206 and Erasmus Placement Mobity. Special tanks to Prof. Stefano Squartini.